 \documentclass[twocolumn,showpacs, floatfix, 
amsmath,nofootinbib,preprintnumbers,letterpaper]{emulateapj}

\usepackage{graphicx}
\usepackage{verbatim}
\usepackage{natbib}

\usepackage{color,comment,aas_macros}

\begin{document}

\title{The synergy between the Dark Energy Survey and the South Pole Telescope}

 \author{Alberto Vallinotto}
 \email{avallinotto@lbl.gov}
 \affiliation{Space Sciences Laboratory, University of California, Berkeley, CA 94720, USA}
 \affiliation{Lawrence Berkeley National Laboratory, Berkeley, CA 94720 USA}

\date{\today}

\begin{abstract}
The Dark Energy Survey (DES) has recently completed the Science Verification phase (SV), collecting data over 150 sq.~deg.~of sky. In this work we analyze to what extent it is beneficial to supplement the analysis of DES data with CMB \textit{lensing} data. We provide forecasts for both DES-SV and for the full survey covering 5000 sq.~deg. We show that data presently available from DES-SV and SPT-SZ would allow a $\sim 8\%$ measurement of the linear galaxy bias in three out of four redshift bins. We further show that a joint analysis of cosmic shear, galaxy density and CMB lensing data allows to break the degeneracy between the shear multiplicative bias, the linear galaxy bias and the normalization of the matter power spectrum. We show that these observables can thus be self calibrated to the percent or sub-percent level, depending on the quality of available data, fraction of overlap of the footprints and priors included in the analysis.
\end{abstract}

\maketitle

\section{Introduction}

Observations of the cosmic microwave background at high redshift and high resolution (ACT\footnote{http://www.physics.princeton.edu/act}, SPT\footnote{http://pole.uchicago.edu}, Planck\footnote{http://www.rssd.esa.int/index.php?project=planck}, PolarBear\footnote{http://bolo.berkeley.edu/polarbear/}) and of galaxies and cosmic shear at low redshift (DES\footnote{http://www.darkenergysurvey.org/}, LSST\footnote{http://www.lsst.org/}, Pan-Starrs\footnote{http://pan-starrs.ifa.hawaii.edu/public/}, Euclid\footnote{http://sci.esa.int/jump.cfm?oid=45403;}, WFIRST\footnote{http://wfirst.gsfc.nasa.gov/}) are taking great strides toward a precise characterization of the universe. While each observational effort targets a specific category of objects/phenomena, a fuller understanding of the universe does arise from conceiving it as a network of interrelated phenomena. On the one hand, galaxy and cosmic shear surveys target luminous objects that grow in the gravitational potential wells sourced by the dark matter overdensities. On the other hand, high resolution CMB \textit{lensing} experiments measure \citep{Smith:2007rg, Das:2011ak, vanEngelen:2012va, Collaboration:2013fk} the \textit{matter} overdensity along a given line of sight integrated all the way to the last scattering surface. Since galaxies populate the scaffolding provided by the dark matter cosmic web, a non-negligible cross-correlation between these ``tracers'' and the CMB lensing signal was expected \citep{Das:2008am} and has recently been detected \citep{Sherwin:2012mr, Bleem:2012gm, Holder:2013hqu, Geach:2013zwa}. Most importantly, because CMB lensing is sensitive \textit{only} to the matter field, such cross-correlations carry fundamental information about the biasing relation existing between the tracers and the density field. Two relevant examples are provided by the \textit{shear multiplicative bias} and by the \textit{linear galaxy bias}.

Crucial to the success of any cosmic shear survey is the constraining of the \textit{multiplicative bias}, a systematic introduced in the shear measurement by the algorithms correcting for atmospheric seeing and for the instrumental distortions of the telescope's point spread function \citep{Heymans:2005rv, Huterer:2005ez, Amara:2007as}. This systematic is particularly insidious because it is completely degenerate with the normalization of the density power spectrum and can lead to a serious degradation in the accuracy of the measured cosmological parameters \citep{Huterer:2005ez, Amara:2007as, Semboloni:2008da}. A first method to constrain it using the impact of lensing on the size and luminosity distributions of galaxies was proposed in Vallinotto et al.~\citep{Vallinotto:2010qm}. Also, Vallinotto \citep{Vallinotto:2011ge} showed how in principle the cross-correlation of CMB lensing and cosmic shear measurements can be used to constrain this systematic.

The \textit{linear galaxy bias} $b(z)$ is the simplest (and crudest) physical observable relating the number overdensity of galaxies $\delta_g(k,z)$ to the overdensity in the dark matter $\delta(k,z)$ through $\delta_g(k,z)\equiv b(z)\delta(k,z)$. While this relation is expected to break down when non-linearities in structure formation start to become significant\footnote{A generalization of the technique proposed here to the case of a scale and redshift dependent bias $b(k, z)$ is straightforward.}, the galaxy bias still carries relevant physical information about the clustering of galaxies on large scales. Constraining the redshift dependence of this bias is crucial for any experiment aiming at probing the nature of gravity through measurements of the growth of structure.

In this work we show how supplementing cosmic shear and galaxy density measurements with cosmological information from CMB lensing can be used to constrain both the galaxy bias and the shear multiplicative bias. These constraints in turn lead to a significant improvement in the constraints that these costly surveys can put on cosmological parameters at no additional cost. While the treatment is completely general, here we present forecasts for the Dark Energy Survey (DES) coupled with present and future measurements of CMB lensing.
DES is a photometric survey measuring weak lensing and galaxy density over 5000 sq.~deg.~in the redshift range $z\in[0.0;1.3]$ \footnote{https://www.darkenergysurvey.org/reports/proposal-standalone.pdf.}. Currently, DES has completed the science verification run (SV) covering 150 sq.~deg.~at full depth.
For CMB lensing we consider two cases: the current run of the South Pole Telescope (SPT-SZ), which has already generated CMB lensing maps for 2500 sq.~deg., and a hypothetical next generation polarization experiment  (CMB-X) with specifications similar to ACT-Pol or a shallower version of SPT-3G and  covering 4000 sq.~deg.
\begin{table}[t]
\begin{center}
\begin{tabular}{c|c|c}
\hline
Parameter & DES + SPT-SZ &  DES + SPT-SZ \\
  & No Planck prior & Planck Prior \\
\hline\hline
$b_0$ & 1.05e-01 & 3.37e-02 \\ 
$b_1$ & 7.92e-02 & 4.02e-02 \\ 
$b_2$ & 7.16e-02 & 5.07e-02 \\ 
$b_3$ & 7.55e-02 & 4.78e-02 \\ 
%(Correct for SPT-SZ)
\hline
\end{tabular}
\caption{Fractional errors on the linear galaxy biases forecasted at $L_{\rm max}=3000$ for DES SV and SPT-SZ.}
\label{Tab:Results_SV1}
\end{center}
\end{table}

\section{Observables} 

Supplementing cosmic shear and galaxy density data with information from CMB lensing should provide a way to constrain the biases characterizing the formers and thus should lead to improved constraints on the cosmological parameters \citep{Vallinotto:2011ge}. We consider three observables: the convergence field extracted from the CMB experiment ($\kappa^{\rm obs}$) using optimal quadratic estimators \citep{Hu:2001kj, *Hirata:2003ka}, the average convergence field measured by the galaxy survey using cosmic shear in the i-th redshift bin ($\bar{\kappa}_i^{\rm obs}$) and the galaxy density field measured by the galaxy survey in the same redshift bin ($\delta_i^{\rm obs}$). We choose the pixelization in Fourier space, so that
\begin{eqnarray}
\kappa^{\rm obs}_{lm} &=& \kappa_{lm}+\kappa^N_{lm},\\
\bar{\kappa}_{i,lm}^{\rm obs} &=& (1+m_i)\,\bar{\kappa}_{i,lm}+\bar{\kappa}^N_{i,lm},\\
\delta_{i,lm}^{\rm obs} &=& \delta_{i,lm}+\delta^N_{i,lm},
\end{eqnarray}
where $m_i$ represents the shear multiplicative bias and the superscript ``N'' denotes the noise contributions. Assuming that weak lensing and galaxy density are measured in $n$ redshift bins, for each set of $\{l,m\}$ there are $2n+1$ observables $\{\kappa^{\rm obs}_{lm}, \bar{\kappa}^{\rm obs}_{1,lm},...,\bar{\kappa}^{\rm obs}_{n,lm}, \delta^{\rm obs}_{1,lm},...,\delta^{\rm obs}_{n,lm}\}$. The total number of observables $N_{\rm tot}$ is thus
\begin{equation}
N_{\rm tot}=a \sum_{l=1}^{l_{\rm max}}(2l+1),
\end{equation}
where $a=2n$ ($a=1$) for the region observed \textrm{only} by the galaxy (CMB lensing) survey and $a=(2n+1)$ for the region where these overlap.

\section{General treatment of correlators} 

Using Limber's approximation, it is straightforward to show that all auto and cross spectra between two of the above physical observables (denoted by $A$ and $B$) take the generic form
\begin{equation}
C_{AB}(l)=\int_0^{\infty}d\chi\frac{g_{A}(\chi)\,g_{B}(\chi)}{\chi^2}\mathcal{P}_{\delta}\left(\frac{l}{\chi},\chi\right),
\end{equation}
where $\chi$ denotes the comoving distance and $\mathcal{P}_{\delta}(k,\chi)$ the matter power spectrum. The $g$ functions encode how each observable is tied to the underlying density field and contributes to the correlation signal. 
The window function for the CMB lensing, weak lensing convergence and galaxy density fields (denoted by respectively by $g_{\kappa}$, $g_{\bar{\kappa},i}$ and $g_{\delta,j}$) are given by
\begin{eqnarray}
g_{\kappa}(\chi)&\equiv&\frac{3\Omega_m H_0^2}{2c^2}\,\frac{D(\chi)D(\chi_{\rm CMB}-\chi)}{D(\chi_{\rm CMB})\,a(\chi)},\\
g_{\bar{\kappa},i}(\chi)&\equiv&\frac{3\Omega_m H_0^2}{2c^2\,a(\chi) \, \bar{\eta}_{i}}\int_{\chi}^{\infty}d\chi'\,
\eta(\chi')\frac{D(\chi)D(\chi'-\chi)}{D(\chi')},\\
g_{\delta,j}(\chi)&\equiv&\eta(\chi)\,b_j\,\Pi(\chi;\chi_j,\chi_{j+1})/\bar{\eta}_j,\\
\bar{\eta}_{i}&\equiv&\int_{0}^{\infty}d\chi\,\eta(\chi)\Pi(\chi;\chi_i,\chi_{i+1}),
\end{eqnarray}
where $\chi_{\rm CMB}$ denotes the comoving distance to the last scattering surface and $\Pi(\chi;\chi_i,\chi_{i+1})$ is a top hat window function for the $i$-th redshift bin stretching from $\chi_i$ to $\chi_{i+1}$. Furthermore, $D(\chi)$ denotes the comoving angular diameter distance, $b_j$ represents the galaxy bias in the $j$-th redshift bin and $\eta(\chi)\equiv d N_g(\chi)/ d\Omega$ is the galaxy number density per unit of solid angle \textit{observed} by the survey at comoving distance $\chi$.

Next, consider the auto and cross-spectra of the noise terms. Because the physical observables are measured with different techniques or by different experiments altogether, it is reasonable to assume the noise cross-spectra to be uncorrelated with respect to one another,
\begin{equation}
\langle \kappa^N_{lm}\,\bar{\kappa}^N_{i,lm}\rangle=\langle\bar{\kappa}^N_{i,lm}\,\delta^N_{j,lm}\rangle=
\langle\delta^N_{j,lm}\,\kappa^N_{lm}\rangle=0\quad\forall i,j.
\end{equation}
\begin{figure}[t]
\includegraphics[width=\columnwidth]{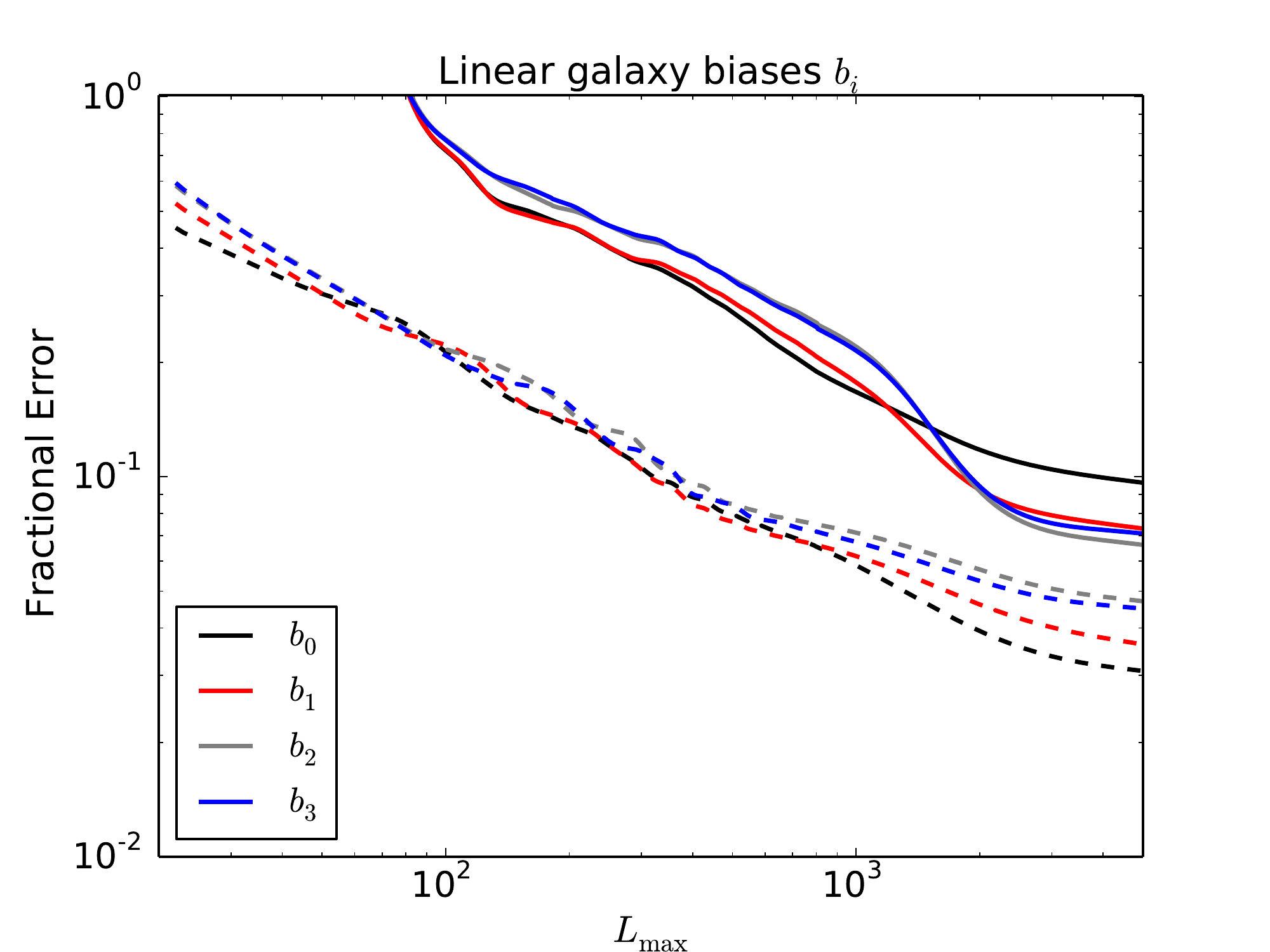}
\caption{\label{Fig:SV1-Biases}Fractional error forecasted for the linear bias measured in each of the four redshift bins $z=0-0.5-0.8-1-1.3$ as a function of the maximum multipole $L_{\rm max}$ included in the analysis. The solid curves show results for DES + SPT-SZ lensing. The dashed curves show the effect of including also a Planck prior on the cosmological parameters.}
\end{figure}
\begin{figure*}[t]
\includegraphics[width=\columnwidth]{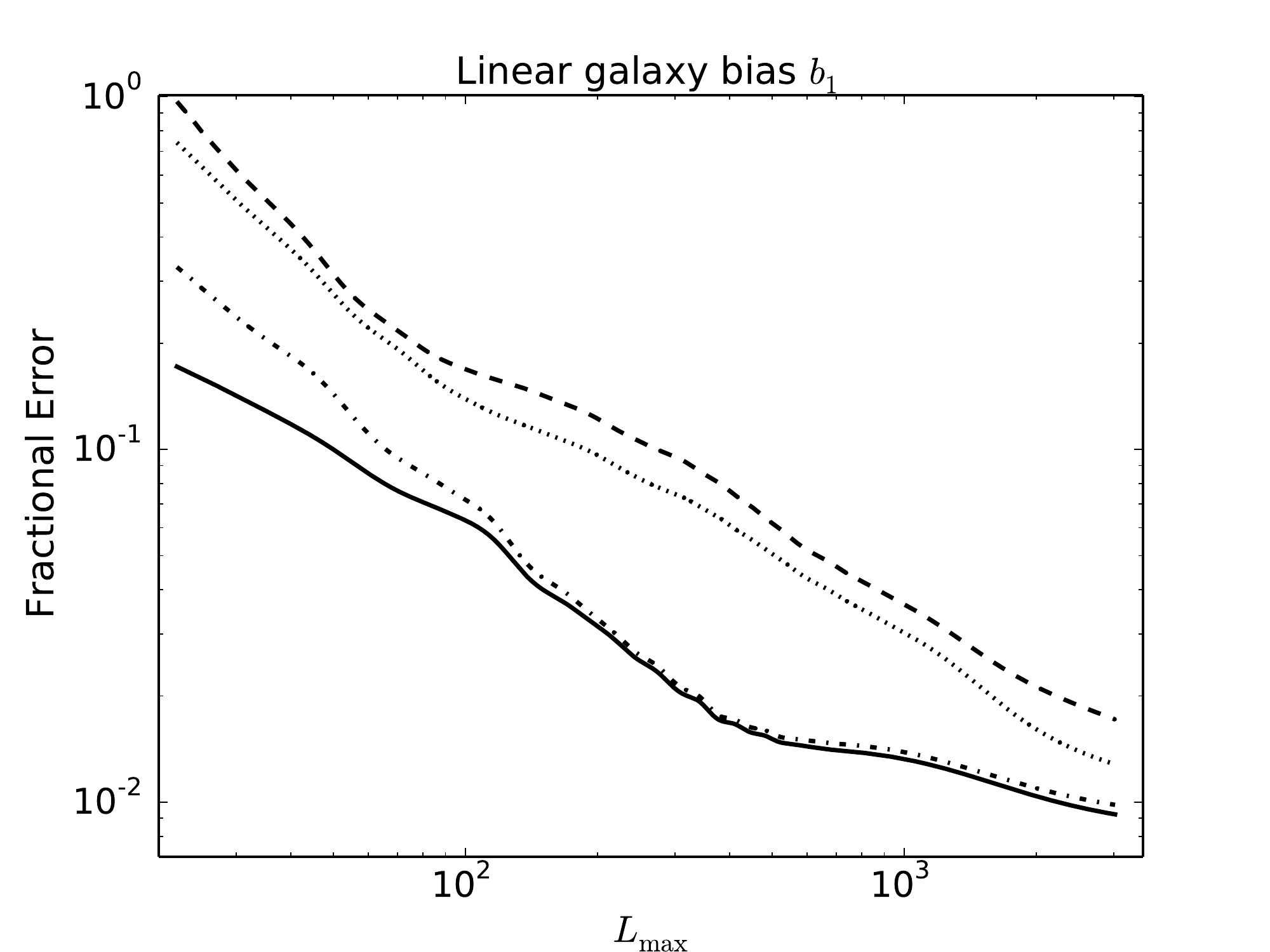}
\includegraphics[width=\columnwidth]{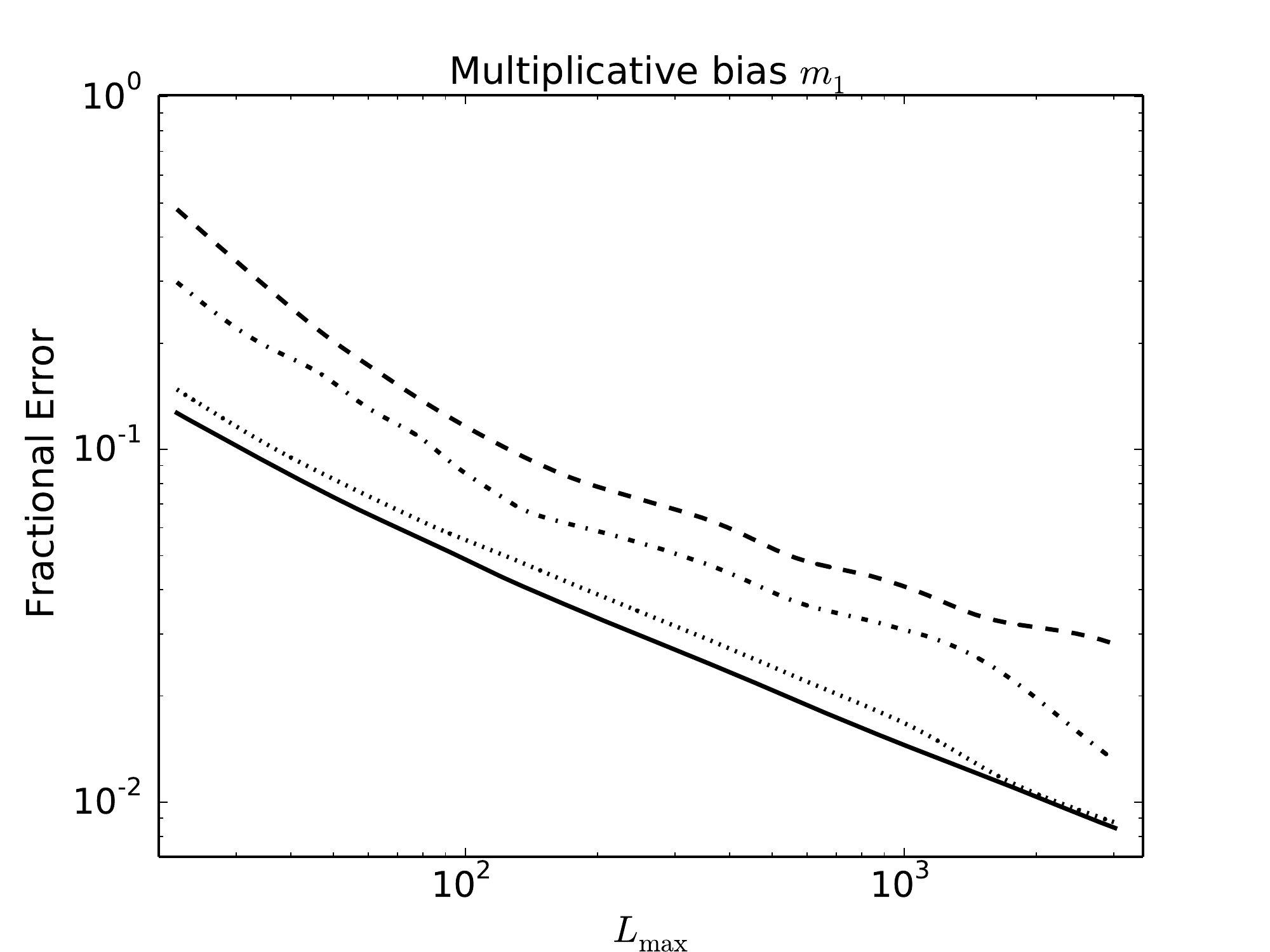}
\caption{\label{Fig:Analyses} Constraints on 
$b_1$ (left panel) and $m_1$ (right panel) 
forecasted for DES + CMB lensing with no footprint overlap (dashed), DES + CMB lensing with no footprint overlap and Planck prior (dot-dashed), DES + CMB lensing with full overlap (dotted), DES + CMB lensing with full overlap and Planck prior (solid) as a function of the maximum multipole $L_{\rm max}$ included in the analysis. The constraints on the other bias parameters closely follow the ones plotted here.}
\end{figure*}
The noise auto spectra depend on the characteristics of the respective experiments. For the observables measured by DES, they are given by
\begin{eqnarray}
C^N_{\bar{\kappa}_i\bar{\kappa}_j}(l)&=&\delta_{ij}\langle\gamma^2_{\rm int}\rangle/\bar{\eta}_{i},\\
C^N_{\delta_{i}\delta_{j}}(l)&=&\delta_{ij}/\bar{\eta}_{i},
\end{eqnarray}
where $\langle\gamma_{\rm int}^2\rangle^{1/2}$ is the rms intrinsic shear. 

To quantify the impact of including cross-correlation information and of overlapping the experiments' footprints, we use the Fisher information matrix 
\begin{eqnarray}
F_{\alpha\beta}=\frac{1}{2}{\rm Tr}\left[C_{,\alpha}C^{-1}C_{,\beta}C^{-1}\right],\label{Eq:Fisher0}
\label{eq:Fisher}
\end{eqnarray}
where $C$ ($C_{,\alpha}$) represent the (derivative with respect to a generic parameter $\alpha$ of the) observables' covariance matrix. To consider different scenarios where the experiments' footprints overlap to different degrees, we proceed as follows. Given a specific configuration, we first quantify the area of the regions where the two experiments footprint do or do not overlap. Then for each distinct region we calculate the Fisher information matrix, including the cross-correlation information of the different observables as appropriate. The Fisher matrices thus obtained are then added together to yield the total Fisher matrix pertaining to that particular experimental configuration.

If including cross-correlation information allows to constrain the biases, it is then important to quantify the impact of this on the measurement of the cosmological parameters. 
The parameters of interest for the analysis are thus the multiplicative and galaxy biases $\{m_i, b_i\}$ for each redshift bin and the cosmological parameters $\{\Omega_m, \Omega_b, h, w_0, N_{\rm eff}, n_s, A_s, \sigma_8 \}$. The forecasts presented next are obtained making the following assumptions:
\begin{itemize}
\item DES: the distribution of sources $d N_g(\chi)/ d\Omega$ is measured directly from DES mock catalogs covering 220 sq.~deg.~with $i\le 24$ magnitude cut. The $dN_g/dz$ distribution peaks around $z\simeq 0.6$, with a number density of weak lensing sources of $\sim 14.7$. We assume the conservative value $\langle\gamma_{\rm int}^2\rangle^{1/2}=0.35$ and four redshift bins: $0-0.5-0.8-1-1.3$. 
\item CMB lensing: we use the noise curves $C^N_{\kappa\kappa}$ calculated for SPT-SZ (CMB-X) assuming CMB maps with noise level of 18 (5) $\mu$K-arcmin.
\item The fiducial cosmology assumed is a flat $\Lambda$CDM cosmology consistent with Planck \citep{Collaboration:2013uq}. The constraints presented for a given parameter are obtained by marginalizing over all the other ones.
\end{itemize}

\section{Forecasts for DES-SV and SPT-SZ}

For DES-SV we assume a footprint of 150 sq.~deg.~over which only galaxy densities (no cosmic shear) are measured at full depth, consistent with the survey's present status. For SPT we assume a 2500 sq.~deg.~footprint completely overlapping DES' and CMB lensing measured according to the telescope's current noise curve. 
The fractional errors forecasted for the linear galaxy bias measured in the four redshift bins are summarized in Tab.~\ref{Tab:Results_SV1} and shown in Fig.~\ref{Fig:SV1-Biases} as a function of the maximum multipole $L_{\rm max}$ included in the analysis. Since $\sigma_8$ and $b_i$ are completely degenerate, a forecast for the constraints on $b_i$ cannot be obtained for DES data alone. The solid curves show the constraints obtained for DES-SV + SPT lensing. The dashed curves show the effect of including also a Planck prior on the cosmological parameters. 
These results show that when analyzed together, the \textit{existing} data from DES-SV and SPT-SZ can lead to a measurement of the linear galaxy bias of the order of $\sim 8\%$ in the last three redshift bins. The inclusion of a Planck prior leads to projected errors of the order of $\sim 5\%$.

\section{Forecasts for DES full survey and CMB-X}

In this case for DES we assume the full 5000 sq.~deg.~footprint, with measurements of both galaxy density and cosmic shear. For CMB-X we use a 4000 sq.~deg.~footprint and the lensing reconstruction noise curves projected for an experiment with a depth of 5 $\mu$K-arcmin.

In Fig.~\ref{Fig:Analyses} we show the constraints on $b_1$ and $m_1$ (the others bias parameters closely following these ones) as a function of $L_{\rm max}$ for four different cases: 
DES + CMB-X lensing with no footprint overlap (dashed), DES + CMB-X lensing with full footprint overlap (dotted), DES + CMB-X lensing with no footprint overlap and Planck prior (dot-dashed) and DES + CMB-X lensing with full footprint overlap and a Planck prior (solid). The projected errors on all the parameters are summarized in Tab.~\ref{Tab:Results_Full}\footnote{In the ``DES only'' case we add a 50\% prior on the completely degenerate set $\{\sigma_8, b_i, m_i\}$ in order to invert the Fisher matrix which would otherwise be singular.}. They show that when information from CMB lensing is included in the analysis, the multiplicative and linear galaxy bias can be constrained to the percent or sub percent level depending on whether the experiments' footprints overlap and whether a prior on the cosmological parameter is added. Also, considering the constraint on the equation of state of dark energy, these results show that, regardless of whether a prior is assumed or not for the cosmological parameters, overlapping the experiments' footprints leads quite generally to a sensible improvement in the error on $w$.
\begin{table}[t]
\begin{center}
\begin{tabular}{c|c|c|c|c|c}
\hline
 & DES &  D+CL & D+CL & D+CL & D+CL \\
 & Only & No ovlp & Full ovlp &  No ovlp & Full ovlp \\
 & & & & Plnk Prior & Plnk Prior \\
\hline\hline
$\sigma_8$ & 2.08e-01 & 7.77e-02& 2.59e-02& 2.74e-02 & 1.92e-02 \\
$\Omega_m$ & 4.04e-02& 3.81e-02& 3.16e-02& 3.05e-03& 2.97e-03\\
$\Omega_b$  & 1.38e-01& 1.22e-01& 1.05e-01& 4.53e-03& 4.51e-03 \\
$N_{\rm eff}$  & 2.09e-01& 1.98e-01& 1.76e-01& 9.22e-02& 7.96e-02 \\
$w$  & 4.47e-02& 4.12e-02& 3.38e-02& 3.03e-02 & 2.23e-02\\
$n_s$  & 2.31e-02& 1.63e-02& 1.02e-02& 2.40e-03 & 2.36e-03\\
$A_s$  & 8.51e-02 & 5.61e-02& 4.29e-02& 1.91e-02 & 1.81e-02\\
$h$  & 6.63e-02& 4.53e-02& 1.59e-02& 1.43e-02 & 1.13e-02\\
$m_0$  & 1.70e-01& 3.51e-02 & 1.96e-02& 2.20e-02 &  1.93e-02\\
$m_1$  & 1.69e-01& 2.81e-02 & 8.78e-03& 1.32e-02 &  8.48e-03\\
$m_2$  & 1.68e-01& 2.71e-02 & 8.19e-03& 1.28e-02&  7.99e-03\\
$m_3$  & 1.68e-01& 2.64e-02 & 7.48e-03& 1.22e-02 &  7.30e-03\\
$b_0$  & 1.67e-01& 1.73e-02 & 1.15e-02& 7.16e-03 &  6.67e-03\\
$b_1$  & 1.67e-01& 1.72e-02 & 1.28e-02& 9.84e-03 &  9.25e-03\\
$b_2$  & 1.67e-01& 1.81e-02 & 1.30e-02& 1.14e-02 &  1.08e-02\\
$b_3$  & 1.67e-01& 1.76e-02 & 1.38e-02& 1.14e-02 &  1.06e-02\\
\hline
\end{tabular}
\end{center}
\caption{Fractional errors on each of the parameters (all the other ones having been marginalized over) estimated at $L_{\rm max}=3000$ for the full DES (D) and CMB-X lensing (CL) surveys.}
\label{Tab:Results_Full}
\end{table}

\section{Discussion}

The key point of all the results presented thus far is the following. For a survey aiming at constraining cosmology through cosmic shear and galaxy density measurements, the set of parameters $\{\sigma_8, b_i, m_i\}$ is completely degenerate. It is then clear that \textit{any} information that can potentially break these degeneracies, whenever added to the analysis, will improve the constraints on the cosmological and bias parameters. In the results presented above, three different kinds of information are at play in breaking these degeneracies: the Planck prior on the cosmological parameters, the cosmological information carried by CMB lensing \textit{alone} and the cross-correlation information (CMB lensing + galaxies and CMB lensing + cosmic shear) arising when the experiments' footprints do overlap. Their effect can be clearly noted by comparing the curves in Fig.~\ref{Fig:Analyses} and considering columns 1-5 of Tab.~\ref{Tab:Results_Full}. First, it is possible to point out that just the inclusion of CMB lensing information, even in a patch of sky \textit{not} overlapping with the one surveyed by the galaxy survey, allows to drastically constrain the galaxy and multiplicative bias to a few percent. Next, considering the dashed curves (DES + CMB-X, no overlap, no prior) as a baseline, Fig.~\ref{Fig:Analyses} shows that multiplicative bias seems to be more sensitive to the overlapping of the footprints while the linear galaxy bias is more sensitive to the prior on cosmological parameters. In particular, it is possible to note that overlapping the footprints allows to significantly constrain the multiplicative bias. Comparison between columns 2 and 3, and between columns 4 and 5 of Tab.~\ref{Tab:Results_Full} shows that these enhanced constraints on the multiplicative bias lead to improvements in the constraints on the cosmological parameters. These improvements compete with and outweigh the reduction of cosmological information arising from the fact that the experiments' footprints are overlapping.

Furthermore, it is necessary to point out the following caveats. First, in the analysis that we carried out we did not include the effect of photometric redshift errors. Since the redshift bins used are quite wide, it is reasonable to expect that photo-z errors will have a limited impact on the conclusions drawn in this work. The investigation of this particular aspect is the focus of a current work \citep{Vallinotto_inprogress}. Second, the results reported in Tab.~\ref{Tab:Results_SV1} and \ref{Tab:Results_Full} are obtained assuming that DES will reach its stated goal of surveying galaxies down to 24 mag. If this goal is not met, the corresponding forecasts are going to degrade accordingly.

Finally, it seems also relevant to point out the following two general facts: first, the constraints on the multiplicative bias forecasted in this work for DES also represent an upper bound for \textit{any} future cosmic shear survey with a density of galaxies comparable or better than DES' (like the LSST) provided that its footprint will also overlap with/include the CMB lensing field. As shown extensively in Tab.~\ref{Tab:Results_Full}, this technique allows to significantly improve the constraints on cosmological parameters \textit{at no additional cost}. Second, since CMB lensing depends only on the distribution of the density field, the cross-correlation of \textit{any} physical observable with it will allow to constrain and extract the biasing relation of the latter. The results reported in this work represent a good example of this latter fact, showing that the actual nature of the biasing relation is not so relevant: while the linear galaxy bias is a physical quantity, the shear multiplicative bias is a systematic. Nonetheless they can both be constrained significantly by a cross-correlation with CMB lensing. It seems therefore possible to conclude by speculating that CMB lensing may provide the ultimate calibration tool for galaxy and cosmic shear surveys, allowing them to constrain their systematics and to self-calibrate their observations to high accuracy.

\textit{Acknowledgements:} A special thanks goes to Gil Holder and Gabrielle Simard, for providing the CMB lensing noise curves as without their help such a detailed work would not have been possible. It is also a pleasure to thank Salman Habib, Katrin Heitmann, Uros Seljak, Eric Linder, Sudeep Das, Scott Dodelson, Hee-Jong Seo and Carlos Cunha for very useful discussions and comments during different stages of this work. Finally, I also thank the Kavli Institute for Cosmological Physics and the Department of Astronomy and Astrophysics at the University of Chicago, Argonne National Laboratory and the Institute for the Early Universe at Ewha Womans University for the hospitality during the various stages of this project. This work has been supported by DOE grant DE-SC-0007867.

%%%%%%%%%%%%%%%%%%%%%%%%%%%%%%%%%%%%%%%%
%%%%%%%%%%%%%%%%%%%%%%%%%%%%%%%%%%%%%%%%
\bibliography{Bibliography}

\begin{thebibliography}{19}
\expandafter\ifx\csname natexlab\endcsname\relax\def\natexlab#1{#1}\fi

\bibitem[{Ade {et~al.}(2013{\natexlab{a}})}]{Collaboration:2013uq}
Ade, P. A.~R., {et~al.} 2013{\natexlab{a}}

\bibitem[{Ade {et~al.}(2013{\natexlab{b}})}]{Collaboration:2013fk}
---. 2013{\natexlab{b}}

\bibitem[{Amara \& Refregier(2008)}]{Amara:2007as}
Amara, A., \& Refregier, A. 2008, Mon. Not. Roy. Astron. Soc., 391, 228

\bibitem[{Bleem {et~al.}(2012)Bleem, van Engelen, Holder, Aird, Armstrong,
  {et~al.}}]{Bleem:2012gm}
Bleem, L., van Engelen, A., Holder, G., Aird, K., Armstrong, R., {et~al.} 2012

\bibitem[{Das {et~al.}(2011)Das, Sherwin, Aguirre, Appel, Bond,
  {et~al.}}]{Das:2011ak}
Das, S., Sherwin, B.~D., Aguirre, P., Appel, J.~W., Bond, J.~R., {et~al.} 2011,
  Phys.Rev.Lett., 107, 021301

\bibitem[{Das \& Spergel(2009)}]{Das:2008am}
Das, S., \& Spergel, D.~N. 2009, Phys.Rev., D79, 043509

\bibitem[{Geach {et~al.}(2013)Geach, Hickox, Bleem, Brodwin, Holder,
  {et~al.}}]{Geach:2013zwa}
Geach, J., Hickox, R., Bleem, L., Brodwin, M., Holder, G., {et~al.} 2013

\bibitem[{Heymans {et~al.}(2006)}]{Heymans:2005rv}
Heymans, C., {et~al.} 2006, Mon. Not. Roy. Astron. Soc., 368, 1323

\bibitem[{Hirata \& Seljak(2003)}]{Hirata:2003ka}
Hirata, C.~M., \& Seljak, U. 2003, Phys.Rev., D68, 083002

\bibitem[{Holder {et~al.}(2013)}]{Holder:2013hqu}
Holder, G., {et~al.} 2013

\bibitem[{Hu \& Okamoto(2002)}]{Hu:2001kj}
Hu, W., \& Okamoto, T. 2002, Astrophys.J., 574, 566

\bibitem[{Huterer {et~al.}(2006)Huterer, Takada, Bernstein, \&
  Jain}]{Huterer:2005ez}
Huterer, D., Takada, M., Bernstein, G., \& Jain, B. 2006, Mon. Not. Roy.
  Astron. Soc., 366, 101

\bibitem[{Semboloni {et~al.}(2008)Semboloni, Tereno, van Waerbeke, \&
  Heymans}]{Semboloni:2008da}
Semboloni, E., Tereno, I., van Waerbeke, L., \& Heymans, C. 2008

\bibitem[{Sherwin {et~al.}(2012)Sherwin, Das, Hajian, Addison, Bond,
  {et~al.}}]{Sherwin:2012mr}
Sherwin, B.~D., Das, S., Hajian, A., Addison, G., Bond, J.~R., {et~al.} 2012,
  Phys.Rev., D86, 083006

\bibitem[{Smith {et~al.}(2007)Smith, Zahn, \& Dore}]{Smith:2007rg}
Smith, K.~M., Zahn, O., \& Dore, O. 2007, Phys.Rev., D76, 043510

\bibitem[{Vallinotto(2012)}]{Vallinotto:2011ge}
Vallinotto, A. 2012, Astrophys.J., 759, 32

\bibitem[{Vallinotto {et~al.}(2011)Vallinotto, Dodelson, \&
  Zhang}]{Vallinotto:2010qm}
Vallinotto, A., Dodelson, S., \& Zhang, P. 2011, Phys.Rev., D84, 103004

\bibitem[{Vallinotto {et~al.}(2013)}]{Vallinotto_inprogress}
Vallinotto, A., {et~al.} 2013, in prep.

\bibitem[{van Engelen {et~al.}(2012)van Engelen, Keisler, Zahn, Aird, Benson,
  {et~al.}}]{vanEngelen:2012va}
van Engelen, A., Keisler, R., Zahn, O., Aird, K., Benson, B., {et~al.} 2012,
  Astrophys.J., 756, 142

\end{thebibliography}

%%%%%%%%%%%%%%%%%%%%%%%%%%%%%%%%%%%%%%%%
%%%%%%%%%%%%%%%%%%%%%%%%%%%%%%%%%%%%%%%%
\end{document}